\documentclass[10pts,showpacs,preprint,aps]{revtex4}
\pdfoutput=1
\usepackage{graphicx}
\usepackage{dcolumn}
\usepackage{bm}
\usepackage{amssymb}
\usepackage{amsmath}
\usepackage{mathrsfs}
\begin{document}
\setcounter{page}{1}
\vskip 2cm
\title
{The Black-Hole radiation in massive gravity}
\author
{Ivan Arraut$^{(1,2)}$}
\affiliation{$^1$Department of Physics, Faculty of Science, Tokyo University of Science,
1-3, Kagurazaka, Shinjuku-ku, Tokyo 162-8601, Japan}
\affiliation{$^2$Department of Physics, Osaka University, Toyonaka, Osaka 560-0043, Japan}

\begin{abstract}
We apply the Bogoliubov transformations in order to connect two different vacuums, one~located at past infinity and another located at future infinity around a black hole inside the scenario of the nonlinear theory of massive gravity. The presence of the extra degrees of freedom changes the behavior of the logarithmic singularity and, as a consequence, the relation between the two Bogoliubov coefficients. This has an effect on the number of particles, or equivalently, on the black hole temperature perceived by observers defining the time arbitrarily.    
\end{abstract}
\maketitle 

\section{Introduction}

\section{Massive Gravity Formulation}   \label{eq:defa}

In the standard formalism of the de-Rham-Gabadadze-Tolley (dRGT) theory of massive gravity, the action is given by the following \cite{dRGT, Kodama}: 
\begin{equation}   \label{eq:b1}
S=\frac{1}{2\kappa^2}\int d^4x\sqrt{-g}(R+m^2U(g,\phi))
\end{equation}
with the effective potential depending on two free parameters by
\begin{equation}   \label{eq:b2}
U(g,\phi)=U_2+\alpha_3 U_3+\alpha_4U_4
\end{equation}
where
\begin{equation}   \label{eq:b3}
U_2=Q^2-Q_2
\end{equation}
\begin{equation}   \label{eq:b4}
U_3=Q^3-3QQ_2+2Q_3
\end{equation}
\begin{equation}   \label{eq:b5}
U_4=Q^4-6Q^2Q_2+8QQ_3+3Q_2^2-6Q_4
\end{equation}
\begin{equation}   \label{eq:b6}
Q=Q_1,\;\;\;\;\;Q_n=Tr(Q^n)^\mu_{\;\;\nu}
\end{equation}
\begin{equation}   \label{eq:b7}
Q^\mu_{\;\;\nu}=\delta^\mu_{\;\;\nu}-M^\mu_{\;\;\nu}
\end{equation}
\begin{equation}   \label{eq:b8}
(M^2)^\mu_{\;\;\nu}=g^{\mu\alpha}f_{\alpha\nu}
\end{equation}
\begin{equation}   \label{eq:b9}
f_{\mu \nu}=\eta_{ab}\partial_\mu\phi^a\partial_\nu\phi^b
\end{equation}

{Here $\phi^a$ are the St\"uckelberg fields, which in general are defined as
\begin{equation}
\phi^a=x^a+\pi^a(r,t)
\end{equation}

However, in this paper, we work in the unitary gauge, such that $\pi^a(r,t)=0$ and $\phi^a=x^a$. Under such conditions, the fiducial metric becomes Minkowskian: $f_{\mu\nu}=\eta_{\mu\nu}$. We note that the metric $f_{\mu\nu}$ is invariant under the Galilean symmetry defined as $\phi^a\to \phi^a+c^a_\mu x^\mu$, with $c^a$ defined as constants \cite{C-dRham}}. We can the compute the field equations as follows \cite{K, Kodama, dRGT}:
\begin{equation}   \label{eq:b10}
{\bf E}_{\mu \nu}=-m^2X_{\mu\nu}
\end{equation}
where
\begin{equation}   \label{eq:b11}
X_{\mu \nu}=\frac{\delta U}{\delta g^{\mu \nu}}-\frac{1}{2}Ug_{\mu \nu}
\end{equation}

Here $f_{\mu \nu}$ is the fiducial metric, and $Q$ is the trace of the matrix $Q^\mu_{\;\;\nu}$. The potential ($U$) defined above is {the only combination able to reproduce a ghost-free theory \cite{{dRGT}}. After variation, the field equations are defined as in Equation (\ref{eq:b10}) (excluding the cosmological constant). Here $E_{\mu \nu}$ is the well-known Einstein tensor obtained from the standard Einstein--Hilbert action (the curvature part in Equation (\ref{eq:b1})), and $X_{\mu \nu}$ has been defined in Equation (\ref{eq:b11})}. 

\section{The St\"uckelberg Trick: The Two Points of View}   \label{eq:Stutwop}

\subsection{St\"uckelberg Fields Appearing in the Dynamical Metric}   \label{eq:herela}

The St\"uckelberg trick as has been formulated  in massive gravity gives us the opportunity of working under two different points of view. The first corresponds to the introduction of the extra degrees of freedom inside the dynamical metric. In such a case, the St\"uckelberg degrees of freedom enter through the metric as follows \cite{K}:
\begin{equation}   \label{eq:gmunu}
g_{\mu\nu}\to {\bf Z}_{\mu\nu}=\frac{\partial Y^\alpha}{\partial x^\mu}\frac{\partial Y^\beta}{\partial x^\nu}g_{\alpha\beta}(Y(x))
\end{equation}
where $Y^\alpha$ contains the information of the extra degrees of freedom. The previous equation resembles the standard gauge transformation in GR. However, it represents in reality the introduction of redundant variables in order to restore the diffeomorphism invariance of the theory. These redundant variables contain in general the information of the extra degrees of freedom of the theory. In fact, the~tensorial object written in Equation (\ref{eq:gmunu}) is gauge invariant, as can be easily proved. The graviton mass term in massive gravity explicitly breaks the diffeomorphism invariance. However, after introducing the St\"uckelberg fields in the form given in Equation (\ref{eq:gmunu}), the massive term is gauge invariant under the following transformation:
\begin{equation}   \label{eq:gt}
g_{\mu\nu}\to\frac{\partial f^\alpha}{\partial x^\mu}\frac{\partial f^\beta}{\partial x^\nu}g_{\alpha\beta}(f(x)), \;\;\;\;\;Y^\mu(x)\to f^{-1}(Y(x))^\mu
\end{equation}

It is easy to verify that under this transformation, Equation (\ref{eq:gmunu}) is gauge invariant, as is demonstrated in \cite{K}. The infinitesimal St\"uckelberg expansion defined as
\begin{equation}   \label{eq:Thisy}
Y^\alpha(x)= x^\alpha+A^\alpha(x)
\end{equation}
provides the following result:
\begin{equation} 
\begin{aligned}
 \label{eq:ob}
{\bf Z}_{\mu\nu}\approx &g_{\mu\nu}+A^\lambda\partial_\lambda g_{\mu\nu}+\partial_\mu A^\alpha g_{\alpha\nu}+\partial_\nu A^\alpha g_{\alpha\mu}+\frac{1}{2}A^\alpha A^\beta\partial_\alpha\partial_\beta g_{\mu\nu}+\partial_\mu A^\alpha\partial_\nu A^\beta g_{\alpha\beta}\\
&+\partial_\mu A^\alpha A^\beta\partial_\beta g_{\alpha\nu}+\partial_\nu A^\alpha A^\beta\partial_\beta g_{\mu\alpha}+...
\end{aligned}
\end{equation}

If we make the infinitesimal expansion $f(x)=x+\zeta(x)$ inside Equation (\ref{eq:gt}) then we obtain the following results 
\begin{equation}
\delta g_{\mu\nu}=\zeta^\lambda\partial_\lambda g_{\mu\nu}+\partial_\mu\zeta^\lambda g_{\lambda\nu}+\partial_{\bf{\nu}}\zeta^\lambda g_{\mu\lambda}
\end{equation}
\begin{equation}
\delta Y(x)= -\zeta^\mu(Y),\;\;\;\;\;\delta A^\mu=-\zeta^\mu-A^\alpha\partial_\alpha\zeta^\mu-\frac{1}{2}A^\alpha A^\beta\partial_\alpha\partial_\beta\zeta^\mu-...
\end{equation}

The $A^\mu$-term corresponds to the Goldstone bosons that at the nonlinear level carry the broken symmetries in massive gravity. It can be verified again that under the previous infinitesimal gauge transformations, Equation (\ref{eq:ob}) gives the result
\begin{equation}
\delta {\bf Z}_{\mu\nu}=0
\end{equation}

In massive gravity, we replace the ordinary perturbation object $h_{\mu\nu}=g_{\mu\nu}-g_{\mu\nu}^{(0)}$ by the object
\begin{equation}   \label{eq:hmunu}
H_{\mu\nu}={\bf Z}_{\mu\nu}-g_{\mu\nu}^{(0)}
\end{equation}
where $g_{\mu\nu}^{(0)}$ is the absolute metric. In this case, it also corresponds to the background metric. Equation~(\ref{eq:hmunu}), when expanded infinitesimally, becomes
\begin{equation}   \label{eq:1}
H_{\mu\nu}=h_{\mu\nu}+\nabla_\mu^{(0)}A_\nu+\nabla_\nu^{(0)}A_\mu
\end{equation}
where the indices for $A_\mu$ are lowered with the background metric. It can be verified that Equation (\ref{eq:1}) has the standard structure of the St\"uckelberg replacement. If the background metric is Minkowskian, then Equation (\ref{eq:1}) can be expanded as
\vspace{12pt}
\begin{equation}   \label{eq:expansion}
H_{\mu\nu}=h_{\mu\nu}+\partial_\mu A_\nu+\partial_\nu A_\mu+\partial_\mu A^\alpha\partial_\nu A_\alpha+...
\end{equation}

In this case, the indices for $A^\mu$ are lowered by using the Minkowskian metric $\eta_{\mu\nu}$. It is standard inside the formulation of the S\"uckelberg trick to introduce the $U(1)$ gauge symmetry transformation by making  the following replacement:
\begin{equation}   \label{eq:repla}
A_\mu\to A_\mu+\partial_\mu\phi
\end{equation}

Then the expansion Equation (\ref{eq:expansion}) takes the form
\begin{equation}
H_{\mu\nu}=h_{\mu\nu}+\partial_\mu A_\nu+\partial_\nu A_\mu+2\partial_\mu\partial_\nu\phi+\partial_\mu A^\alpha\partial_\nu A_\alpha+\partial_\mu A^\alpha\partial_\nu\partial_\alpha\phi+\partial_\mu\partial^\alpha\phi\partial_\nu A_\alpha+\partial_\mu\partial^\alpha\phi\partial_\nu\partial_\alpha\phi
\end{equation}
and the gauge transformation becomes
\begin{equation}
\begin{aligned}
\delta h_{\mu\nu}&=\partial_\mu\zeta_\nu+\partial_\nu\zeta_\mu+\pounds_\zeta h_{\mu\nu}\\
\delta A_\mu&=\partial_\mu\Lambda-\zeta_\mu-A^\alpha\partial_\alpha\zeta_\mu-\frac{1}{2}A^\alpha A^\beta\partial_\alpha\partial_\beta\zeta_\mu-...\\
\delta\phi&=-\Lambda
\end{aligned}
\end{equation}

More details about the St\"uckelberg method and its extensions can be found in \cite{Stu,Stu1,Stu2,Stu3}.

\subsection{St\"uckelberg Fields Appearing in the Fiducial Metric}

Another way of exploring the St\"uckelberg formalism is by introducing the fields inside the fiducial metric. In such a case, the dynamical metric transforms covariantly as usual \cite{K}. The idea is to make the following replacement:
\begin{equation}   \label{eq:Miaujunior}
g_{\mu\nu}^{(0)}\to f_{\mu\nu}=g_{\alpha\beta}^{(0)}\partial_\mu Y^\alpha\partial_\nu Y^\beta
\end{equation}
where the St\"uckelberg fields transform as scalars under diffeomorphism transformations:
\begin{equation}
Y^\alpha(x)\to Y^\alpha(f(x))
\end{equation}
or infinitesimally
\begin{equation}
\delta Y^\alpha(x)=\zeta^\beta\partial_\beta Y^\alpha(x)
\end{equation}

Then the fiducial metric behaves as a tensor under diffeomorphism transformations. The~perturbation $h_{\mu\nu}=g_{\mu\nu}-g_{\mu\nu}^{(0)}$ can then be replaced by
\begin{equation}
H_{\mu\nu}=g_{\mu\nu}-f_{\mu\nu}
\end{equation}
with $f_{\mu\nu}$ defined as in Equation (\ref{eq:Miaujunior}). Using again the transformation Equation (\ref{eq:Thisy}), together with $h_{\mu\nu}=g_{\mu\nu}-g_{\mu\nu}^{(0)}$, we obtain
\begin{equation}
H_{\mu\nu}=h_{\mu\nu}+g_{\nu\alpha}^{(0)}\partial_\mu A^\alpha+g_{\mu\alpha}^{(0)}\partial_\nu A^\alpha-g_{\alpha\beta}^{(0)}\partial_\mu A^\alpha\partial_\nu A^\beta
\end{equation}

Again, using the replacement Equation (\ref{eq:repla}), we obtain
\begin{equation}
H_{\mu\nu}=h_{\mu\nu}+\partial_\mu A_\nu+\partial_\nu A_\mu+2\partial_\mu\partial_\nu\phi+\partial_\mu A^\alpha\partial_\nu A_\alpha+\partial_\mu A^\alpha\partial_\nu\partial_\alpha\phi+\partial_\mu\partial^\alpha\phi\partial_\nu A_\alpha\\
+\partial_\mu\partial^\alpha\phi\partial_\nu\partial_\alpha\phi
\end{equation}

Under infinitesimal gauge transformations, we have
\begin{equation}
\begin{aligned}
\delta h_{\mu\nu}&=\partial_\mu\zeta_\nu+\partial_\nu\zeta_\mu+\pounds_\zeta h_{\mu\nu}\\
\delta A_\mu&=\partial_\mu\Lambda-\zeta_\mu+\zeta^\nu\partial_\nu A_\mu\\
\delta\phi&=-\Lambda
\end{aligned}
\end{equation}

Another way of introducing the St\"uckelberg fields can be found in \cite{stu2,stu3}.

\section{The Schwarzschild de-Sitter Solution in dRGT} \label{eq:Final1}

In \cite{Kodama}, the S-dS solution was derived for two different cases. The first corresponds to the family of solutions satisfying the condition $\beta=\alpha^2$, where $\beta$ and $\alpha$ correspond to the two free parameters of the theory. In such a case, the St\"uckelberg function $T_0(r,t)$ becomes arbitrary. The~second corresponds to the family of solutions with two free parameters satisfying the condition $\beta\leq\alpha^2$ with the St\"uckelberg function constrained. The generic black hole solution is given explicitly as
\begin{equation}
ds^2={Z_{tt}dt^2+Z_{rr}S_0^2dr^2+Z_{rt}(drdt+dtdr)}+S_0^2r^2d\Omega_2^2
\end{equation}
where

\begin{equation}   \label{eq:drgt metric}
{ Z_{tt}=-f(S_0r)(\partial_tT_0(r,t))^2,\\
Z_{rr}=-f(S_0r)(\partial_rT_0(r,t))^2+\dfrac{1}{f(S_0r)},\\
Z_{tr}=-f(S_0r)\partial_tT_0(r,t)\partial_rT_0(r,t)}
\end{equation}

Here we define $f(S_0r)=1-\frac{2GM}{S_0r}-\frac{1}{3}\Lambda S_0^2 r^2$. In this previous solution, all the degrees of freedom are inside the dynamical metric. Because we are working in unitary gauge, then the fiducial metric in this case is the Minkowski metric:
\begin{equation}
f_{\mu\nu}dx^\mu dx^\nu=-dt^2+dr^2+r^2(d\theta^2+r^2sin^2\theta)
\end{equation}
where $S_0=\dfrac{\alpha}{\alpha+1}$ is just a scale factor depending on the free parameters of the theory, namely, the two appearing in the potential term defined in Equation (\ref{eq:b2}) \cite{Kodama}. The St\"uckleberg fields take the standard form defined in \cite{Kodama}. The solution Equation (\ref{eq:drgt metric}) can be equivalently written in a generic form:
\begin{equation}   \label{eq:drgt metric2}
ds^2=-f(S_0r)dT_0(r,t)^2+\frac{S_0^2dr^2}{f(S_0r)}+S_0^2r^2d\Omega^2
\end{equation}
where $T_0(r,t)$ corresponds to the St\"uckelberg function. In this case however, this function contains the information of the extra degrees of freedom in agreement with the formulation of Section \ref{eq:herela}. {In~other words, $T_0(r,t)$ is not the ordinary time coordinate}. In fact, the metric Equation (\ref{eq:drgt metric2}), with the definitions of Equation (\ref{eq:drgt metric}), is gauge invariant under the transformations defined in Equation (\ref{eq:gt}) with $T_0(r,t)=f(Y(r,t))$. The functions $Y^\alpha$ are explicitly given by
\begin{equation}   \label{The result}
T_0=Y^0=S_0t+A(r, t),\;\;\;\;\;Y^r=S_0r
\end{equation}
by using the same notation as in \cite{Kodama} and the conventions of Section \ref{eq:herela}. 

\section{The Particle Creation Process}   \label{eq:Hawking}

The particle creation process in black holes is a consequence of the fact that in curved spacetimes, the concept of a vacuum is not absolute. The vacuum can only be defined locally. Two different vacuums are connected through the Bogoliubov transformations. Because the concept of a vacuum is ambiguous for the case of curved spacetimes, then the concept of particles will be equivalently ambiguous. The~definition of a vacuum is connected to the way in which the time-like Killing vector is defined locally. The definition of this vector is related to the way in which the positive frequencies for the different modes are taken. The first derivation of the black hole radiation was done in \cite{Hawking,Hawking2}. Here we follow the same arguments, and then we give their extension in order to include the effects of the extra degrees of freedom in the theory of massive gravity. We divide the analysis into two different cases. The first corresponds to the case in which the observers take the time in agreement with $T_0(r,t)$. This case is exactly the same as in GR. The second corresponds to that in which the observers define the time arbitrarily.
 
\subsection{Observers Defining the Time in Agreement with $T_0(r,t)$}

{For simplicity, in this analysis, we omit the cosmological constant term $\Lambda$, such that we can focus on the role of the extra degrees of freedom in the black hole radiation, as is perceived by observers located at large scales}. {However, this particular case} will correspond to a review of the situation described inside the scenario of GR. This is the case because the observers, moving such that their time coordinate is equivalent to $T_0(r,t)$, will perceive the same physics as in GR ({they will not be able to perceive the effect of the extra degrees of freedom}). Here we consider the standard expansion for a scalar field in terms of positive and negative frequencies:
\begin{equation}   \label{expansionlalala}
\phi=\sum_i\left(f_i\hat{a}_i+\bar{f}_i\hat{a}_i^+\right)
\end{equation}

\begin{figure}
	\centering
		\includegraphics[width=10cm, height=8cm]{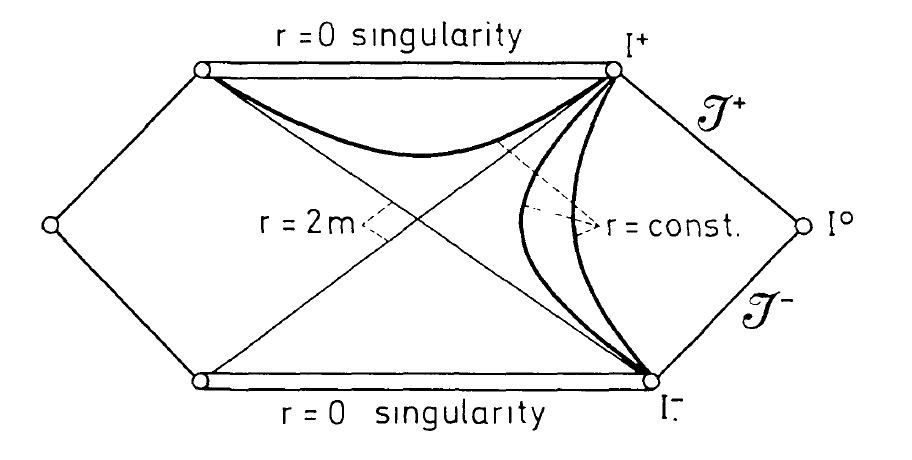}
	\caption{The Penrose diagram for the Schwarzschild geometry in general relativity (GR) as is shown in \cite{Hartle}. In massive gravity, the same diagram is valid if we express the black hole solutions in terms of the St\"uckelberg functions, defining then the time coordinate in agreement with $T_0(r,t)$. We note that for arbitrary time definitions, the same diagram will not be valid even if we have  a zero cosmological constant.}
	\label{fig:momoko}
\end{figure}

Here we take the functions $f_i$ as a complete family forming an orthonormal set over past infinity defined as $\mathscr{I}^-$ in the Penrose diagram shown in the figure \ref{fig:momoko}. The orthonormality condition is defined~as   
\begin{equation}   \label{productlala}
\frac{1}{2}i\int_S\left(f_i\bar{f}_{j;a}-\bar{f}_jf_{i;a}\right)d\Sigma^a=\delta_{ij}
\end{equation} 
\vspace{-12pt}

The integration is done over the surface $S$, which in this case corresponds to the surface defined by $\mathscr{I}^-$. The semicolons denote the derivative with respect to the variable changing along $\mathscr{I}^-$, and the upper bar denotes complex conjugation. The previous equation defines the inner product for the orthonormal basis defined by $f_i$ and $\bar{f}_i$. This is in fact the conserved inner product of the Klein--Gordon equation. In the expansion Equation (\ref{expansionlalala}), the operators $\hat{a}$ and $\hat{a}^+$ are the annihilation and creation operators for particles at past null infinity $\mathscr{I}^-$. Particles at past null infinity are ingoing particles. We note that in agreement with the definition of Equation (\ref{productlala}), the functions $f_i$ would have positive frequency only with respect to the affine parameter along past null infinity $\mathscr{I}^-$. The expansion Equation (\ref{expansionlalala}) for the scalar field is defined in terms of the data at past null infinity. However, we can also expand the scalar field in terms of the functions, defining their data at future null infinity $\mathscr{I}^+$, and the data at the event horizon. Then we can define the scalar field as

\begin{equation}   \label{expansionlala2} 
\phi=\sum_i\left(p_i\hat{b}_i+\bar{p}_i\hat{b}_i^+ +q_i\hat{c}_i+\bar{q}_i\hat{c}_i^+\right)
\end{equation}  

Here the functions $p_i$ define an orthonormal set over future null infinity $\mathscr{I}^+$, and then they will have positive frequencies only with respect to the affine parameter along $\mathscr{I}^+$. The sets at future infinity are outgoing modes, and they have zero Cauchy data at the event horizon. Then the operators $\hat{b}_i$ and $\hat{b}_i^+$ are the annihilation and creation operators for the particles on $\mathscr{I}^+$ (for outgoing particles). On the other hand, the functions $q_i$ have zero Cauchy data at future null infinity. They~form a complete orthonormal set along the future event horizon. The exact definition of the modes $q_i$ is not relevant. The interpretation of the operators $\hat{c}$ and $\hat{c}^+$ is unknown at this level but still irrelevant. At~this point, it becomes important to define the relations between the modes $f_i$ and the modes $p_i$, as~well as the relations between the corresponding annihilation and creation operators. In order to keep the canonical commutation relations unchanged when the fields are quantized, namely,    
\begin{equation}
[\hat{a}_i, \hat{a}_j^+]=[\hat{b}_i, \hat{b}_j^+]=i\delta_{ij}
\end{equation}
then we have to connect the fields and operators through the Bogoliubov transformations as follows:
\begin{equation}   \label{Bogolibov}
p_i=\sum_j\left(\alpha_{ij}f_j+\beta_{ij}\bar{f}_j\right)
\end{equation}  
for the functions and
\begin{equation}
\hat{b}_i=\sum_j\left(\bar{\alpha}_{ij}\hat{a}_j-\bar{\beta}_{ij}\hat{a}_j^+\right)
\end{equation} 
for the operators. We note that here the relations for the creation operators can be found by taking the adjoint operation over the annihilation operator. In order to find the Bogoliubov coefficients, we have to define the modes at past null infinity $\mathscr{I}^-$ ($f_i$), and we also have to define the modes over $\mathscr{I}^+$ ($p_i$). These correspond to the ingoing and outgoing modes respectively, and in spherical coordinates they are 
\begin{equation}   \label{Phase2la}
\begin{aligned}
f_{\omega', l, m}&=(2\pi)^{-1/2}r^{-1}(\omega')^{-1/2}F_{\omega'}(r)e^{i\omega'V}Y_{l,m}(\theta, \phi)\\
p_{\omega, l, m}&=(2\pi)^{-1/2}r^{-1}(\omega)^{-1/2}P_{\omega}(r)e^{i\omega U}Y_{l,m}(\theta, \phi)
\end{aligned}
\end{equation}

Here $U(r, T_0(r,t))$ and $V(r, T_0(r,t))$ are the St\"uckelberg function{\bf s} expressed in terms of advanced and retarded coordinates. They are defined here as
\begin{equation}   \label{stulight}
\begin{aligned}
U(r, T_0(r,t))&=T_0(r,t)+S_0r+2MLog\left\vert \frac{S_0r}{2M}-1 \right\vert=u+A(r,u)\\
V(r, T_0(r,t))&=T_0(r,t)-S_0r-2MLog\left\vert \frac{S_0r}{2M}-1 \right\vert=v+A(r,v)
\end{aligned}
\end{equation}      
taking into account Equation (\ref{The result}). We note that here the nontrivial part of the St\"uckelberg function is expressed as a function of the advanced and retarded coordinates for each case. Having~defined the advanced and retarded St\"uckelberg functions, we have to define the relation between them. From the definition of the St\"uckelberg trick given in Equation (\ref{eq:gmunu}), which resembles a standard coordinate transformation from the GR perspective, we conclude that the Penrose diagrams of the standard GR will be valid as far as we use the St\"uckelberg functions defined appropriately. Near~the event horizons, it is always expected that the St\"uckelberg functions converge to ordinary coordinates, that is, $U(r, T_0(r,t))\approx u$ and $V(r, T_0(r,t))\approx v$, because $A(r,t)\to0$ for strong gravitational fields. Far away from the horizons, the causal structure of the spacetime might change if we express the results in terms of standard coordinates $u$ and $v$. This point is important, as later we see that for this reason the Hawking radiation perceived by observers located at large scales changes when they define their local time arbitrarily. We can assume, without loss of generality, that the spacetime is asymptotically flat when it is expressed in terms of the St\"uckelberg functions. Then we can keep the relation between the advanced and retarded St\"uckelberg functions in the same way as we keep it for the related coordinates in GR for the asymptotically flat case. Then the surfaces of constant phase $\omega U(r, T_0(r,t))$ are 
\begin{equation}   \label{eq:uv}
\omega U=-\frac{\omega}{\kappa_{eff}}\left(log (V_0-V)-log D-log C\right)
\end{equation}
     
This is just the same relation inside GR, which is expected, as we have remarked that the observers defining the time in agreement with the St\"uckelberg function will perceive gravity as GR and as a consequence will perceive the standard black hole radiation. Having then this relation, we~can then express the mode $p_{\omega}^{(2)}$ in terms of the function $V(r, T_0(r,t))$ by using Equations (\ref{Phase2la}) and (\ref{eq:uv}). Here we ignore the angular components represented by the Harmonic functions, which we assume to be normalized in a standard way. Then we can define
\begin{equation}   \label{argument}
p_\omega^{(2)}\approx (2\pi)^{-1/2}r^{-1}(\omega)^{-1/2}P_{\omega}^{-}\left(\frac{V_0-V}{CD}\right)^{-i\frac{\omega}{\kappa_{eff}}}
\end{equation}
 
Here $p_\omega^{(2)}$ represents the portion of $p_\omega$ that can enter the black hole and that is not scattered by the body. {This solution is only valid for $V_0-V$ small and positive. These modes can enter the body because its effective frequency will be very high, obeying then the geometric optics. This means that we can ignore the effects of the potential barrier in the Schwarzschild metric, as it will only scatter low-frequency modes. More details about this argument can be found in \cite{Hawking,Hawking2}.} By taking into account the Bogoliubov relations defined in Equation (\ref{Bogolibov}), we can use the Fourier transformation in order to find the Bogoliubov coefficients. This is done by taking into account the definition of orthonormality introduced in Equation (\ref{productlala}) and {if we take} as conjugate variables under the Fourier transformation {the quantities} $\omega'$ and $V$. Without any surprise, here the results are
\begin{equation}   \label{alphabeta}
\begin{aligned}
\beta_{\omega, \omega'}^{(2)}&\approx -i\alpha_{\omega, (-\omega')}^{(2)}\\
\alpha_{\omega, \omega'}^{(2)}&\approx (2\pi)^{-1}P_{\omega}^{-}(CD)^{i\frac{\omega}{\kappa_{eff}}}e^{i(\omega-\omega')V_0}\left(\frac{\omega'}{\omega}\right)^{1/2}\Gamma\left(1-\frac{i\omega}{\kappa_{eff}}\right)(-i\omega')^{-1+i\frac{\omega}{\kappa_{eff}}}
\end{aligned}
\end{equation}

In addition, the relation between $\alpha^{(2)}$ and $\beta^{(2)}$ is defined by the way in which the logarithmic singularity, given by the term $(-i\omega')^{-1+i\frac{\omega}{\kappa_{eff}}}$, is avoided when we try to connect the two coefficients by analytical extension \cite{Hawking,Hawking2}. The final result is
\begin{equation}   \label{aaaa}
\vert\alpha^{(2)}_{\omega, \omega'}\vert=e^{\frac{\pi\omega}{\kappa_{eff}}}\vert\beta_{\omega, \omega'}^{(2)}\vert
\end{equation} 
which is well known. In what follows, we consider the case in which the observers define the time arbitrarily, which is the interesting case for our purposes. We note that here we have defined the surface gravity as $\kappa_{eff}$, which we consider as the effective surface gravity perceived by an observer defining the time in agreement with $T_0(r,t)$. Later we see that the observers defining the time coordinate arbitrarily perceive a different temperature {for the black hole}.

\subsection{Observers Defining the Time Arbitrarily}

For this case, the results are different with respect to the GR case. The difference comes from the presence of the term $A(r, u)$ defined in Equation (\ref{stulight}). The presence of the term $A(r, v)$ is irrelevant, as we see soon. We note that $A(r, u)$ can be expressed in terms of $v$ if we take into account the well-known relation between advanced and retarded coordinates in the Penrose diagram. The relation between the St\"uckelberg function $U(r, T_0(r,t))$ and the affine parameter, which helps to connect the future event horizon and the surfaces of constant $U(r, T_0(r,t))$, is 
\begin{equation}   \label{epsilon}
\epsilon=Ce^{-\kappa U(r,T_0)}
\end{equation} 

This is just an extension of the well-known relation obtained in the scenario of GR. The difference is that in this case, $U(r,T_0(r,t))=u+A(r, u)$. We note that the previous relation is obtained when we parallel transport the family of null vectors satisfying the condition $l^an_a=-1$. Here $l^a$ is a null vector parallel to the future event horizon, and $n^a$ is a future directed null vector, directed inwards. Then the vector $-\epsilon n^a$ connects the future event horizon with the surface of constant phase $U(r, T_0(r,t))$. We note that in the neighborhood of the event horizon, $U(r, T_0(r,t))\approx u$, because $A(r, u)\to 0$ for strong gravitational fields. However, very far from the future event horizon, we have $A(r, u)\neq0$, and this is the case in which we are located at past null infinity $\mathscr{I}^-$. This is a true event if we are located near the past event horizon, which we take here as the surfaces of constant $V_0\approx v_0$. With this change, we~find that for observers defining the time arbitrarily, the surfaces of constant phase are defined by the relation
\begin{equation}   
\omega u=-\frac{\omega}{\kappa}\left(log \epsilon-log C\right)-\omega A(r,u)
\end{equation} 

From here we can find that the surfaces of constant phase $\omega u$ are defined as
\begin{equation}   \label{therelation}
\omega u=-\frac{\omega}{\kappa}\left(log (v_0-v-A(r,v))-log D-log C-log (e^{\kappa A(r, u)})\right)
\end{equation}

Here we can see explicitly the distortion of time produced by the extra degrees of freedom of the theory through the terms $A(r, v)$ and $A(r, u)$. If we want to express $A(r, u)$ as a function of $v$ instead of $u$, then we need to use Equation (\ref{therelation}) inside this function. This will give us a result of the~form 
\begin{equation}
A(r, u)=A\left(r,\left\{log \left(\frac{v_0-v-A(r, v)}{CDe^{\kappa A(r, u)}}\right)^{-\frac{1}{\kappa}}\right\}\right)
\end{equation}  

We note that in this equation, $A(r, u)$ appears again on the right-hand side. This means that the surfaces of constant phase represented by Equation (\ref{therelation}) have an iterative solution. Here however we simplify the result by considering that the iteration is not necessary and that on the right-hand side of the previous equation, the approximation $\kappa A(r, u)\to 0$ is valid. This simplification is done with the purpose of finding compact results, but the method developed here is general. We can now replace Equation (\ref{therelation}) inside Equation (\ref{Phase2la}), thus obtaining
\begin{equation}   \label{pomega}
p_\omega^{(2)}\backsim (2\pi)^{-1/2}\omega^{-1/2}r^{-1}P_\omega^-\left(\frac{v_0-v-A(r,v)}{CD e^{\kappa A(r,u)}}\right)^{-\frac{i\omega}{\kappa}}
\end{equation} 

If we want to obtain the Bogoliubov coefficients, again we need to use the Fourier transformation by taking the variables $\omega'$ and $v$ as {the} conjugate {variables}. Then we have to solve the integral 
\begin{equation}    \label{pomegaalpha}
\begin{aligned}
\alpha_{\omega, \omega'}^{(2)}\approx& -(2\pi)^{-1}P_{\omega}^{-}(CD)^{i\frac{\omega}{\kappa}}\left(\frac{\omega'}{\omega}\right)^{1/2}\int \left(v_0-v-A(r, v)\right)^{-i\frac{\omega}{\kappa}}e^{i\omega A(r, u)}\\
&\times\left(1+\partial_v A(r, v)\right)e^{-i\omega'(v+A(r, v))}dv
\end{aligned}
\end{equation}

We note that this previous result is obtained after using the orthonormality condition Equation~(\ref{productlala}), applied to Equation (\ref{Bogolibov}) and taking into account Equation (\ref{pomega}). The expansion Equation (\ref{Bogolibov}) can be expressed as
\begin{equation}
p_\omega=\int_0^\infty d\omega'\left(\alpha_{\omega, \omega'}f_{\omega'}+\beta_{\omega, \omega'}\bar{f}_{\omega'}\right)
\end{equation} 
by replacing the sum by the integral operation, taking into account that the frequency is a continuous variable. The application of the orthonormality condition requires the evaluation of the derivative over $\bar{f}_\omega'$, here defined as
\begin{equation}
\bar{f}_{\omega';v}=-i(1+\partial_vA(r,v))(2\pi)^{-1/2}r^{-1}(\omega')^{1/2}F_{\omega'}(r)e^{-i\omega'(v+A(r,v))}
\end{equation} 
       
By multiplying this result with the mode in Equation (\ref{pomega}), we then obtain Equation (\ref{pomegaalpha}) after integration over the variable $v$. We note that the integration depends on how the function $A(r, u)$ is expressed as a function of $v$. We know that the relation between $u$ and $v$ is given by Equation (\ref{epsilon}). Without loss of generality, we can assume that $A(r, u)$ is a polynomial expansion of $u$ given by
\vspace{12pt}
\begin{equation}   \label{Aplo}
A(r, u)\approx \sum_{n=0}^\infty a_nu^n
\end{equation}  
 
{We note that this functional dependence is general enough but is not the only possibility for the functional dependence of $A(r,u)$}. The possible radial dependence of this function is irrelevant as it does not contribute to the integration in Equation (\ref{pomegaalpha}). If we introduce the relation Equation (\ref{therelation}), we~then obtain
\begin{equation}
A(r, u)\approx \sum_{n=0}^\infty a_n\left(-\frac{1}{\kappa}Log\left(\frac{v_0-v-A(r, v)}{DC}\right)\right)^n
\end{equation}

We note that the solution for this equation is iterative. However, in order to develop an example for what happens when the observers define the time arbitrarily, we have ignored the iterative process by taking $e^{\kappa A(r,u)}\to 1$, as we have explained previously.    
 
\subsubsection{Specific Example of the Modification of the Temperature Perceived by Observers Defining the Time Arbitrarily}

In order to develop an example for what happens when the observers define the time arbitrarily, we have selected some specific functional dependence of $A(r, u)$. We have selected a polynomial expansion with respect to $u$. This has the advantage that we can factorize the terms in the integral in an easier way. Then for example, by assuming $\kappa A(r, u)<<1$, we obtain in Equation (\ref{pomegaalpha}) the following~result:
\begin{equation}
\begin{aligned}    \label{pomegaalpha22}
\alpha_{\omega, \omega'}^{(2)}\approx& -(2\pi)^{-1}P_{\omega}^{-}(CD)^{i\omega\sum_n\frac{na_n}{\kappa^n}}\left(\frac{\omega'}{\omega}\right)^{1/2}\int \left(v_0-v-A(r, v)\right)^{-i\omega\sum_n\frac{na_n}{\kappa^n}}\\
&\times\left(1+\partial_v A(r, v)\right)e^{-i\omega'(v+A(r, v))}dv
\end{aligned}
\end{equation}

Here the linear term in the expansion ($n=1$) has absorbed the original factor. If we make the substitution $j=v_0-v-A(r, v)$, the integral to be evaluated in Equation (\ref{pomegaalpha22}) becomes
\begin{equation}   \label{pomegaalpha221}
\alpha_{\omega, \omega'}^{(2)}\approx (2\pi)^{-1}P_{\omega}^{-}(CD)^{i\omega\sum_n\frac{na_n}{\kappa^n}}e^{-i\omega'v_0}\left(\frac{\omega'}{\omega}\right)^{1/2}\int j^{-i\omega\sum_n\frac{na_n}{\kappa^n}}
e^{i\omega'j}dj
\end{equation}
 
In order to obtain the Gamma function form, we need to make the following substitution:
\begin{equation}
x=-i\omega'j
\end{equation}

Then the previous expression becomes

\begin{equation}    \label{pomegaalpha222}
\alpha_{\omega, \omega'}^{(2)}\approx (2\pi)^{-1}P_{\omega}^{-}(CD)^{i\omega\sum_n\dfrac{na_n}{\kappa^n}}e^{-i\omega'v_0}\left(\dfrac{\omega'}{\omega}\right)^{1/2}\left(\int x^{-i\omega\sum_n\dfrac{na_n}{\kappa^n}}
e^{-x}dx\right)\left(-i\omega'\right)^{-1+i\omega\sum_n\dfrac{na_n}{\kappa^n}}
\end{equation}

The integral {inside the} parentheses corresponds to a Gamma function, and thus we obtain the result
\begin{equation}   \label{alphabeta2}
 \begin{aligned}
\beta_{\omega, \omega'}^{(2)}&\approx -i\alpha_{\omega, (-\omega')}^{(2)}\\
\alpha_{\omega, \omega'}^{(2)}&\approx (2\pi)^{-1}P_{\omega}^{-}(CD)^{i\omega\sum_n\dfrac{na_n}{\kappa^n}}e^{i(\omega-\omega')v_0}\left(\dfrac{\omega'}{\omega}\right)^{1/2}\Gamma\left(1-i\omega\sum_n\dfrac{na_n}{\kappa^n}\right)(-i\omega')^{-1+i\omega\sum_n\dfrac{na_n}{\kappa^n}}
 \end{aligned}
\end{equation}	

If we want to obtain $\alpha^{(2)}$ from $\beta^{(2)}$ by analytically extending $\omega'$ through the logarithmic singularity, we obtain the modified relation
\begin{equation}   \label{mamamia}
\vert\alpha^{(2)}_{\omega, \omega'}\vert=e^{\pi\omega\sum_n\frac{na_n}{\kappa^n}}\vert\beta_{\omega, \omega'}^{(2)}\vert
\end{equation} 

We note that here we have assumed some explicit dependence for the function $A(r, u)$. The relation between the coefficients might change depending on the behavior of this function. We note also that the function $A(r, v)$ is irrelevant for the purposes of calculation because it disappears in the substitution of variables. Equation (\ref{mamamia}) is just equivalent to Equation (\ref{aaaa}) but shows the decomposition of the effective surface gravity $\kappa_{eff}$, which the observers defining the time in agreement with $T_0(r,t)$ perceive, in terms of the surface gravity $\kappa$, which observers defining the time arbitrarily perceive. A direct comparison of the mentioned results gives
\begin{equation}   \label{kappaeffective}
\kappa_{eff}=\left(\sum_n\frac{na_n}{\kappa^n}\right)^{-1}=\frac{1}{4GM}
\end{equation}   

Here $\kappa_{eff}$ is the surface gravity for the observers defining the {time in agreement with} $ T_0(r,t)$. On~the other hand, $\kappa$ is the surface gravity for observers defining an arbitrary time. The previous expression shows that if $\kappa_{eff}=1/4GM$, as in the standard case, then the surface gravity $\kappa$ is obtained from the solution of a polynomial equation defined in agreement with Equation (\ref{kappaeffective}). The~order of the polynomial expansion depends on how the function $A(r, u)$ is defined. The definition for this function will change for different observers defining the time in a different way. The larger the deviation of the coordinate time $t$ with respect to the time defined by the St\"uckelberg function $T_0(r,t)$ as it is perceived by the observers, the larger the contribution coming from the function $A(r, u)$ will be, and as a consequence, the deviations of the black hole temperature perceived by an observer will be larger. {Equation (\ref{kappaeffective}) can change if the functional dependence defined in Equation~(\ref{Aplo}) for the function $A(r, u)$ changes. In more general situations, the result will not be as simple as that obtained in Equation (\ref{kappaeffective})}. 

\subsubsection{Discussion of the Results: The Concepts of Time}

{The previous results were based on the way in which the observers define their time locally. In this paper, we have defined two possible coordinate times, namely, the ordinary time coordinate $t\neq T_0(r,t)$ and the St\"uckelberg function $t=T_0(r,t)$. In general, $t\neq T_0(r,t)$, but for some observers it can happen that $t=T_0(r,t)$. These special observers will not perceive the effects of the graviton mass. For them, $A(r,t)=0$ in Equation (\ref{The result}). Selecting an arbitrary time $t\neq T_0(r,t)$ or the St\"uckelberg function $t=T_0(r,t)$ as a time coordinate will depend on how each observer defines the relation between the time coordinate and the proper time. The relation between the proper time and the coordinate time is defined by the symmetry under time translations. In GR, we have a unique definition of a constant, defined as a conserved quantity appearing when there is symmetry under time translations. This is related to the definition of the time-like Killing vector. In massive gravity however, in general it is not possible to define the same symmetry under time translations in a conventional way. It is possible to demonstrate however, that we can define a conserved quantity in the direction of the St\"uckelberg function $T_0(r,t)$ as has been explained in \cite{Mysuper,Mysuper1,Mysuper2,PRD}. Then we can define a Killing vector in the direction of $T_0(r,t)$. It turns out that the relation between the proper time and the St\"uckelberg function in massive gravity is equivalent to the relation between the proper time and the ordinary coordinate time $t$ in GR. However, in general, the relation between the proper time and an arbitrary coordinate time ($t\neq T_0(r,t)$) in massive gravity is different to the unique relation defined in GR. Then when we are talking about observers defining the time arbitrarily, we are talking about those observers defining an arbitrary relation between their proper time and the coordinate time $t$. On the other hand, when we talk about observers defining the time in agreement with $T_0(r,t)$, we are talking about those observers defining the relation between the proper time and the coordinate time in agreement with GR, but inside the theory of massive gravity. The results found in this paper represent a first step in the unification of the concepts of Hawking radiation and the Unruh effect \cite{Unruh} from a different perspective. From the results reported here, we can think in a possible generalization of them in order to explain, in a unified way, both effects, namely, Hawking radiation and the Unruh effect, as distortions of the notion of the time coordinate. We consider distortions of the time coordinate, those able to create ambiguities in the concept of a vacuum. Important results proving the link between the Hawking radiation and the Unruh effect have been explored in \cite{New1,New2}.} 

\section{Conclusions}   \label{conclusions}

In this paper, we have evaluated the black hole temperature for the spherically {symmetric} case inside the scenario of the dRGT massive gravity. We have used the method of the Bogoliubov transformations, and {we have taken} into account the relations between vacuums defined at future null infinity and past null infinity in agreement with the Penrose diagram. The Penrose diagram in massive gravity is the same as in the case of GR, where we express the solutions in terms of the St\"uckelberg functions. When expressed in terms of the ordinary coordinates, deviations of the causal structure with respect to the GR case are expected. Here we introduce these possible deviations by considering the behavior of the function $A(r, u)$, which represents the deviation {of the St\"uckelberg function} with respect {to} the ordinary {definition of} time. The observers defining the time in agreement with $T_0(r,t)$, {or in other words, defining the relation between the proper time $\tau$ and the coordinate time $t$ as a relation between the proper time and the St\"uckelberg function $T_0(r,t)$}, will not perceive the contributions coming from {the function $A(r, u)$}; they will perceive the standard result obtained in GR, here called $\kappa_{eff}$. This quantity can be expanded {as a series expansion} with respect to the surface gravity ($\kappa$) perceived by an observer defining the time arbitrarily. The difference between $\kappa_{eff}$ and $\kappa$ makes sense, as the causal structure of the spacetime changes at large scales for the observers defining the time arbitrarily as a result of the presence of the function $A(r, u)$. {Defining the time arbitrarily is equivalent to defining an arbitrary relation between the proper time and the coordinate time $t\neq T_0(r,t)$}. {In addition, we wish to remark that if we want to obtain the surface gravity for an observer defining the time arbitrarily ($t\neq T_0(r,t)$), then we have to solve a polynomial equation for $\kappa$, as is shown in Equation (\ref{kappaeffective}). Then the result is not trivial at all. If the functional behavior of $A(r, u)$ is different, this~changes the relation between $\kappa$ and $\kappa_{eff}$, as well as the relation between the two Bogoliubov coefficients for the case of observers defining the time arbitrarily.} The methods developed in this paper can be repeated with other theories of massive gravity, such as, for example, those breaking Lorentz symmetry~\cite{Rubakov,Rubakov2}. {We also expect these methods to be a first step in finding a novel way of unifying the Hawking radiation effect and the Unruh effect. Other methods explaining the common origin of the Hawking radiation with the Unruh effect have been analyzed in \cite{New1,New2}.} Alternative~methods for calculating the black hole temperature in massive gravity theories have been given in \cite{Mysuper,Mysuper1,Mysuper2}. {It is also important to remark that the result obtained in \cite{PRD} suggests that the effective mass perceived by an observer located at large scales in massive gravity depends on how the observers define the time coordinate locally. Although the result obtained in \cite{PRD} was not oriented to the analysis of black hole radiation, it~is evidently connected with the black hole temperature perceived by the observers in massive gravity. However, the notion of effective mass is not enough in order to find a relation between the surface gravity $\kappa$ and the effective gravity $\kappa_{eff}$, unless we are able to define the functional relation between the mass defined by an observer taking the time as $t=T_0(r,t)$ and the observer defining the time as $t\neq T_0(r,t)$, which was not possible to develop in \cite{PRD}. Then the Bogoliubov method is more general; it extracts the relevant physics of the system, and it is the cleanest way of deriving the statistics of black body radiation when we know the relation between the two coefficients~\cite{Hawking,Hawking2}.} .\\\\

{\bf Acknowledgement}
I. A was supported by the JSPS Post-Doctoral fellow for oversea researchers.

\end{document}